\documentclass[aps,reprint,notitlepage,superscriptaddress,showpacs,showkeys]{revtex4-1}
\usepackage{latexsym,amssymb,amsfonts,mathrsfs,amsmath,bm}
\usepackage{graphicx}
\usepackage[usenames,dvipsnames]{xcolor}
\usepackage{soul}
\usepackage[linktocpage,colorlinks=true,linkcolor=blue,citecolor=blue,breaklinks=true,urlcolor=blue]{hyperref}
\usepackage[detect-all]{siunitx} 
\usepackage{sidecap}

\definecolor{darkgreen}{rgb}{0,0.5,0.0}

\newcommand\al{\textit{et~al.}\ }

\newcommand\rt{\textit{run-and-tumble}}


\begin{document}

\title{\emph{E.coli} ``super-contaminates'' narrow ducts fostered by broad run-time distribution}

\author{Nuris Figueroa-Morales}
\affiliation{Laboratoire de Physique et M\'ecanique des Milieux H\'et\'erog\`enes, PMMH, ESPCI Paris, PSL University, CNRS, Sorbonne Universit\'e, Univ Paris Diderot, 10, rue Vauquelin, 75005 Paris, France}
\affiliation{Department of Biomedical Engineering, The Pennsylvania State University, PA 16802, USA}

\author{Aramis Rivera} 
\affiliation{Zeolites Engineering Lab, IMRE, University of Havana, 10400 Havana, Cuba}

\author{Rodrigo Soto}
\affiliation{Departamento de F\'\i sica, FCFM, Universidad de Chile, Santiago, Chile}

\author{Anke Lindner}
\affiliation{Laboratoire de Physique et M\'ecanique des Milieux H\'et\'erog\`enes, PMMH, ESPCI Paris, PSL University, CNRS, Sorbonne Universit\'e, Univ Paris Diderot, 10, rue Vauquelin, 75005 Paris, France}

\author{Ernesto Altshuler}
\affiliation{Group of Complex Systems and Statistical Physics, Physics Faculty, University of Havana, 10400 Havana, Cuba}

\author{\'Eric Cl\'ement}
\email{eric.clement@upmc.fr}
\affiliation{Laboratoire de Physique et M\'ecanique des Milieux H\'et\'erog\`enes, PMMH, ESPCI Paris, PSL University, CNRS, Sorbonne Universit\'e, Univ Paris Diderot, 10, rue Vauquelin, 75005 Paris, France}

\date{\today}

\begin{abstract}

One striking feature of bacterial motion is their ability to swim upstream along corners and crevices, by leveraging hydrodynamic interactions. 
This motion through anatomic ducts or medical devices might be at the origin of serious infections. 
However, it remains unclear how bacteria can maintain persistent upstream motion while exhibiting \rt~dynamics.
Here we demonstrate that \textit{E. coli} can travel upstream in microfluidic devices over distances of 15 millimeters in times as short as 15 minutes. 
Using a stochastic model relating the run times to the time bacteria spend on surfaces, we quantitatively reproduce the evolution of the contamination profiles when considering a broad distribution of run times. 
Interestingly, the experimental data cannot be reproduced using the usually accepted exponential distribution of run times. Our study demonstrates that the \rt~statistics determine macroscopic bacterial transport properties. 
This effect, that we name ``super-contamination'', could explain the fast onset of some life-threatening medical emergencies. 

\end{abstract}

\maketitle

\section*{Introduction}

Bacteria live in a wide variety of natural environments in which fluid flow is present, including the capillary networks of animals and plants and porous soils \cite{Valdes-Parada2009,Duchesne2010}. Upstream bacterial infections often occur in ducts where liquids are oscillating or flowing in one direction, such as in the human urinary tract and medical catheters \cite{wright2005uropathogenic, Dohnt2011, Kim2012}. Understanding the upstream motility of bacteria in such confined scenarios is crucial to prevent infections or control microbial soil pollution \cite{Rusconi2015}.

An extended motility mechanism in bacteria in open environments is the well known \rt\ dynamics \cite{Berg2004book}. In this strategy the cells moves in a series of straight paths with quick reorientations of the swimming direction, resulting in 3D random walks \cite{Berg2004book}.
Solid surfaces modify the bacterial dynamics, introducing hydrodynamic interactions that lead to surface accumulation and circular trajectories \cite{Berke2008,Lauga2006}.
Increased complexity emerges in shear flows, from the interplay between the flow and confining surfaces and the bacterial structures (fore-aft asymmetry and chiral flagella) \cite{Kaya2012, Marcos2012, mathijssen2018oscillatory}. At low shear rates bacteria can migrate upstream close to the surfaces and the edges of the bounding structures \cite{Hill2007, Kaya2012, Altshuler2013, Figueroa2013, Figueroa2015, mathijssen2018oscillatory}. High shear rates, on the other hand, produce an erosive detachment \cite{Figueroa2015} from the surfaces. The overall transport in a confined channel is then built on diverse contributions: 
downstream advection in the bulk and, depending on the flow velocity, upstream and/or downstream motion along the crevices and close to the surfaces.

%
%

In spite of its potential importance, to our knowledge, typical distances for upstream swimming have not been previously determined. 
Hydrodynamic interactions between the swimmer and the surface are related to the velocity and geometry of the swimmer \cite{Berke2008}. For a bacterium undergoing tumble, the hydrodynamic interactions will be decreased and the erosion process enhanced.
On this logic, the statistics of tumbles and runs should critically determine the upstream bacterial contamination inside microchannels.

For \textit{Escherichia coli}, the run time distribution was reported to follow a single-time Poisson process related to the rotational switching of the flagellar motor \cite{berg1972chemotaxis}. More recently, direct measurements on flagellar motors show heavy-tailed distributions of rotation times stemming from the intrinsic noise in the chemotactic signaling \cite{korobkova2004molecular}. Experimental works highlight the existence of very persistent trajectories in swimming bacteria \cite{wu2006collective, Figueroa_Thesis, figueroa20183d}, possibly connected to heavy tailed distributed run times. 
However, most theoretical or numerical studies use a Poisson process to model the microscopic stochasticity of the kinematics \cite{ReviewActiveMatter2013}. 
The influence of this stochastic processes on the macroscopic bacterial transport remains an open question \cite{matthaus2011origin, sneddon2012stochastic, figueroa20183d}.

Here, by means of a video-scanning technique, we study the upstream migration of {\it E. coli} with single-bacteria resolution up to macroscopic distances in excess of \SI{15}{\milli\meter}. 
We find that the contamination takes place in the form of a front of invading bacteria, moving upstream along surfaces and corners, with a front velocity almost independent of the perfusion fluid velocity --we call it ``super-contamination''.
We found this process to be related to the existence of very long run times. 
When the time bacteria spend close to surfaces is linked to the run times and the duration of runs is modeled by a {\it broad} distribution, we find a quantitative agreement with the experiments. However, exponentially distributed run times fail to even qualitatively describe this process. This effect may explain, for example, fast infections in the urinary tract or through medical catheters.

\begin{figure}
\centering      \includegraphics[width=0.75\linewidth]{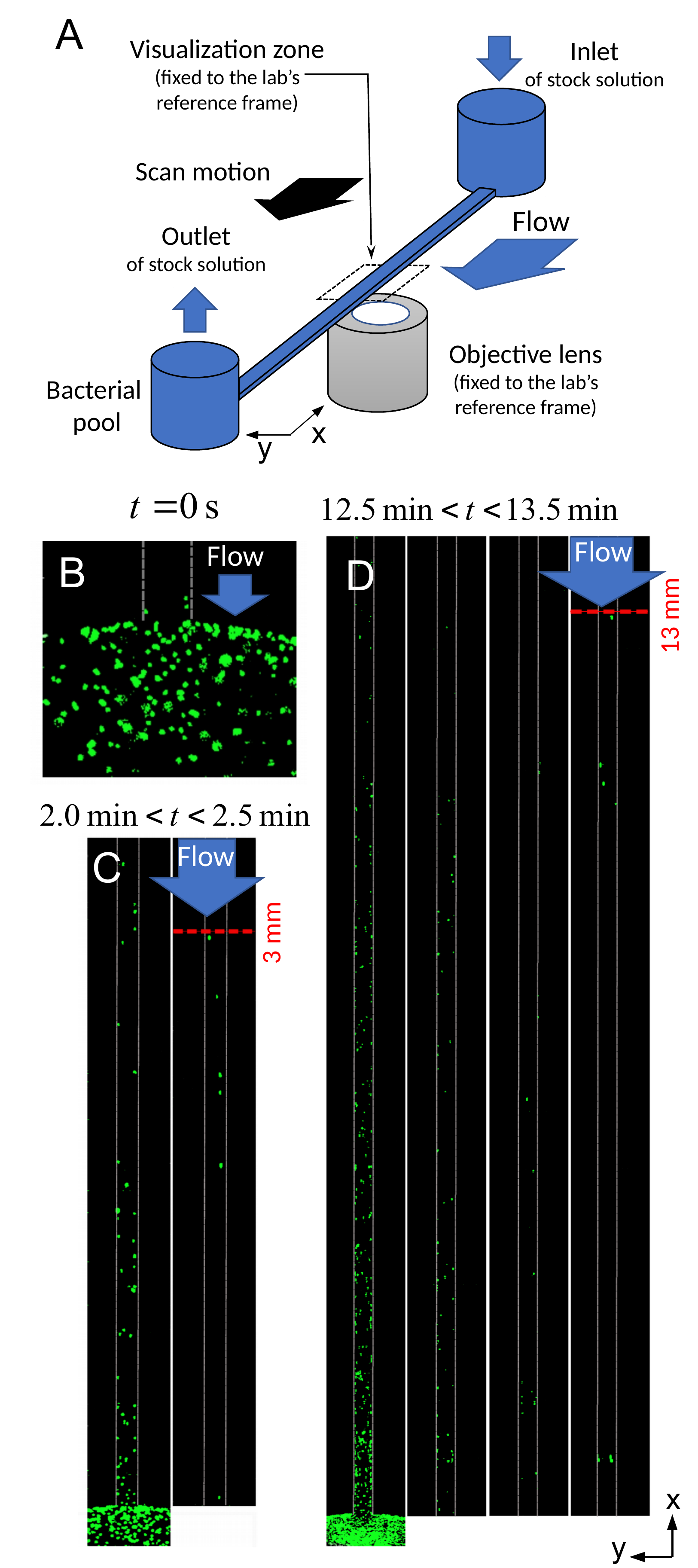}
 \caption{Visualizing upstream super-contamination. (A) Sketch of the microfluidic device. The black arrow indicates the direction of motion of the setup during a scan, while the lens stays in place. (B) Image of swimmers, represented by green spots, near the exit of the bacterial reservoir at the beginning of one contamination experiment. (C) Reconstruction of the channel using images from a scan performed between \SI{2}{\minute} and \SI{2.5}{\minute} from the beginning of the experiment: the pioneer swimmer has reached a distance of \SI{3}{\milli\meter} from the bacterial reservoir. (D) An analogous reconstruction, taken from a scan performed from \SI{12}{\minute} to \SI{13.5}{\minute}: the pioneer swimmer has reached a distance of \SI{13}{\milli\meter}. The flow velocity at the center of the channel was \SI{80}{\micro\meter/\second}. As a size reference, the width of the channel is $w = \SI{40}{\micro\meter}$.}
 \label{fig:scan_montage_vertical}
\end{figure}

\section*{Experimental results}

The experimental setup, as sketched in Fig.~\ref{fig:scan_montage_vertical} (A), consists of a few-mm-long PDMS rectangular channel of width $w = \SI{40}{\micro\meter}$, height $h = \SI{11}{\micro\meter}$, and length $L=\SI{15}{\milli\meter}$, glued on top of a PDMS-covered glass slide. Its extremes end up in two cylindrical reservoirs of $\SI{1}{\milli\meter}$ diameter, connected to tubing that allow fluids to circulate using a gravity flow. 

At the beginning of the experiment, the outlet reservoir is filled with {\it E. coli} bacteria (strain RP437 expressing green fluorescent protein). 
For every experiment, the average bacterial velocity in the reservoir $V_\text{b}$ is measured.
By injecting a sufficiently high flow rate of the fluid without bacteria through the inlet, bacteria are kept in the outlet reservoir, not yet invading the channel. See Materials and Methods for details on the procedure. Then, at $t=0$, a controlled flow of the bacteria-free liquid is established from the inlet to the outlet and the system is video-recorded. Bacteria start swimming upstream along the channel. The flow rate in the channel is determined by tracking passive latex beads suspended in the perfusion fluid. The maximum flow velocity measured in the center of the channel is represented as $V_\text{f}$. 
Since our experimental conditions avoid the presence of chemical gradients, our experiment is different from those of references \cite{saragosti2011directional} and \cite{Emonet2018spatial}.
 
In order to visualize individual bacteria over the macroscopic extent of the channel, which is several millimeters long starting from the left reservoir, the channel is translated at constant velocity $V_\text{s} = \SI{150}{\micro\meter/\second}$ along the $-x$ direction while a video is taken through an inverted microscope fixed on the laboratory reference frame (Fig.~\ref{fig:scan_montage_vertical}(A)). The channel is scanned several times during a single experiment at a fixed flow rate. 
Supplementary Movie S1 shows a sequence of three scans starting at different times during the same experiment.

Fig.~\ref{fig:scan_montage_vertical}(B), (C), and (D) show a combination of images taken during scans starting at different time points. The total time interval associated with one image spans from the beginning of the scan, to the moment of visualization of the farthest bacteria from the reservoir, which we call ``pioneers''. There, we visualize all the bacteria along the channel length, independently of their vertical position. The multiple side-by-side images in panels (C) and (D) are actually end-to-end in the physical system.
Note that this channel reconstruction does not constitute a snapshot, since different segments display the conditions at different instants. However, the pictures clearly demonstrate the arrival of bacteria as far as \SI{13}{\milli\meter} from the bacterial reservoir in \SI{13.5}{\minute}. This constitutes direct evidence for the ability of bacteria to {\it swim upstream along macroscopic distances in a short time interval}.

\begin{figure}
\centering      \includegraphics[width=0.8\linewidth]{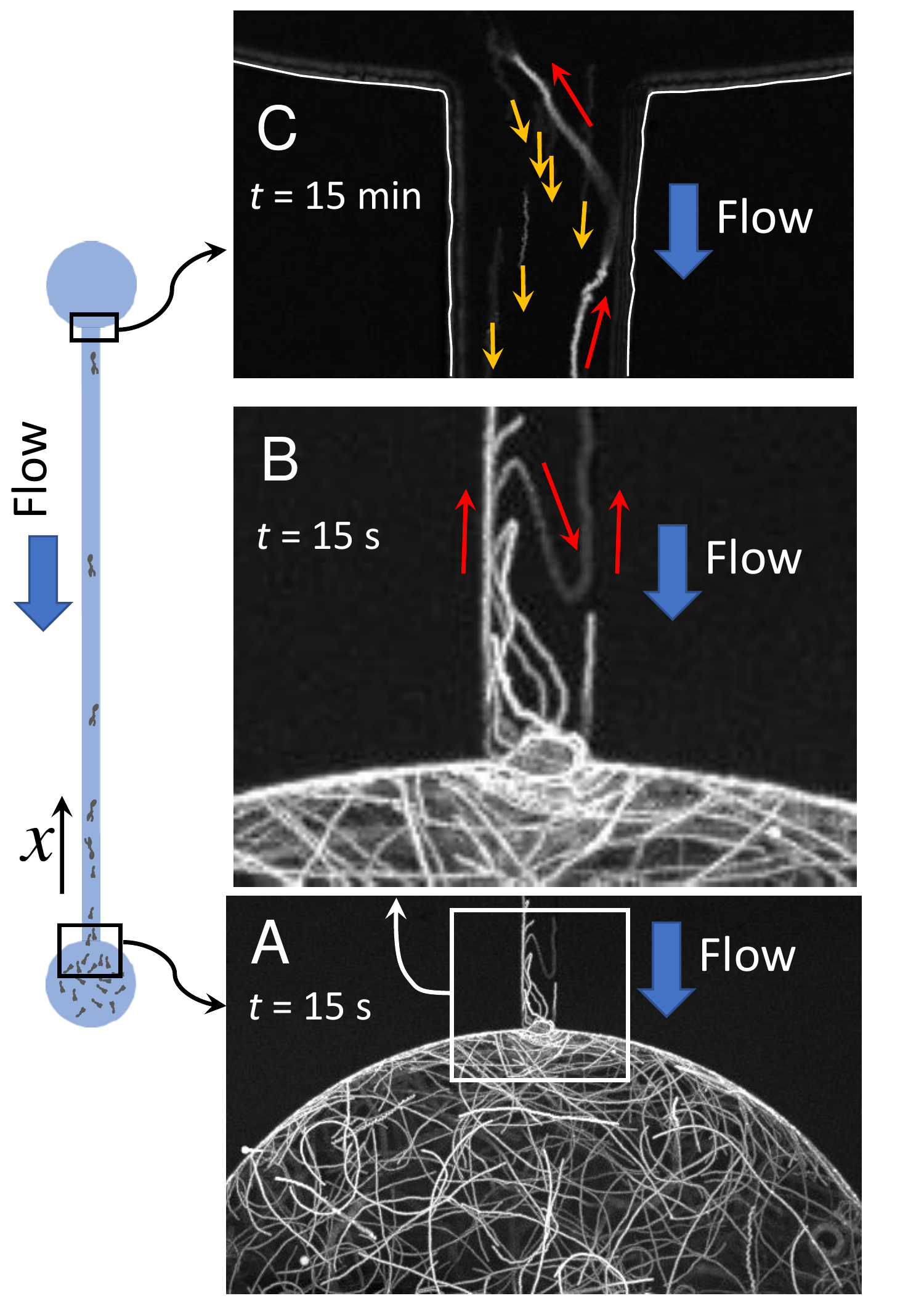}
 \caption{ Bacterial trajectories are represented by the superposition of photograms on fixed positions, taken at intervals of \SI{1/30}{\second}  from \SI{13}{\second} and  \SI{9}{\second} long videos corresponding to (A-B) and (C), respectively. (A) Bacterial trajectories near the entrance of the channel from the bacterial reservoir, \SI{15}{\second} after starting the contamination experiment. (B) Zoom at the entrance of the channel: red arrows indicate the path of one bacterium that first moves upstream along the left wall, then detaches from it, and then re-attaches to the right wall, continuing its upstream motion. (C) A zoom at the end of the channel: red arrows indicate the upstream trajectory of a bacterium that has reached the opposite extreme of the channel after swimming for a distance bigger than \SI{15}{\milli\meter} in \SI{15}{\minute}; the orange arrows indicate the paths of ``inactive'' beads moving downstream (the borders of the channel have been over-drawn in white for clarity). The flow velocity at the center of the channel was of \SI{5}{\micro\meter/\second}. As a reference, the width of the channel is $w = \SI{40}{\micro\meter}$.}
 \label{fig:Trajectories}
\end{figure}

In Fig.~\ref{fig:Trajectories}(A) we display details of the trajectories of some bacteria near the outlet (the bacterial reservoir). Panel (B) is a zoom illustrating how the flow (approximately \SI{5}{\micro\meter/\second} at the center of the channel) induces the bacteria to concentrate near the entrance of the channel. This is associated with bacterial attaching and detaching from the walls of the reservoir and moving progressively towards the channel, a phenomenon previously reported using a funnel geometry~\cite{Altshuler2013}. This densification at the outlet creates favorable conditions to prime an efficient contamination process in the channel, as it will induce higher chances for the bacteria to get inside the microchannel via upstream swimming close to surfaces, but mostly along the channel edges. This last phenomenon is illustrated by the vertical trajectories in the picture. The red arrows indicate a typical contaminating trajectory: after swimming upstream along the left wall of the channel, the bacterium is detached from it, then it is advected downstream and re-attaches to the right wall of the channel, continuing its upstream motion. After many events like this, our persistent swimmer eventually reaches the opposite extreme of the channel. That event is illustrated in Fig.~\ref{fig:Trajectories} (C), where the bacterial trajectory is pointed out by red arrows, while the orange arrows indicate the motion of a few latex beads used to measure the flux.


\begin{figure*}
    \centering    \includegraphics[width=0.8\linewidth]{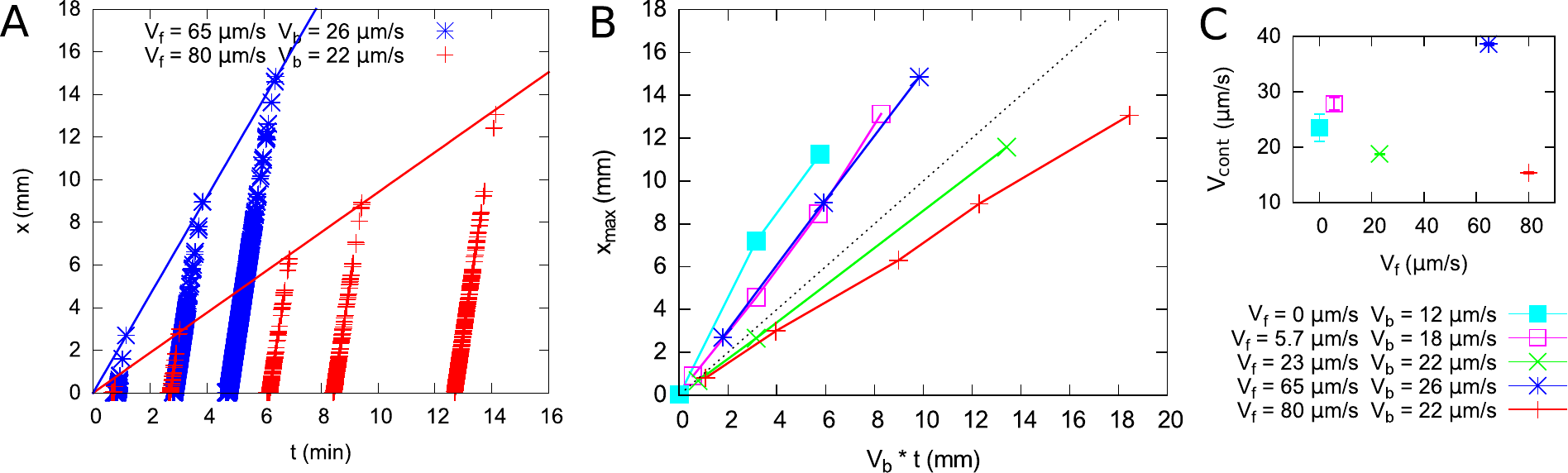}
    \caption{
Quantification of the upstream contamination. (A) Scanning position (dots) for two experiments with different flow velocities. The scanning stops at the farthest bacteria (pioneers). The straight lines show the advance of the contamination pioneers, with a slope that gives the contamination velocity $V_\text{cont}$.
(B) Positions of the pioneers as a function of $V_\text{b}t$.
The dotted line is the curve $y=x$.  
(C)  $V_\text{cont}$ as a function of $V_\text{f}$ for different experiments.
}
    \label{fig:pioneers}
\end{figure*}

\begin{figure*}
\centering      \includegraphics[width=\linewidth]{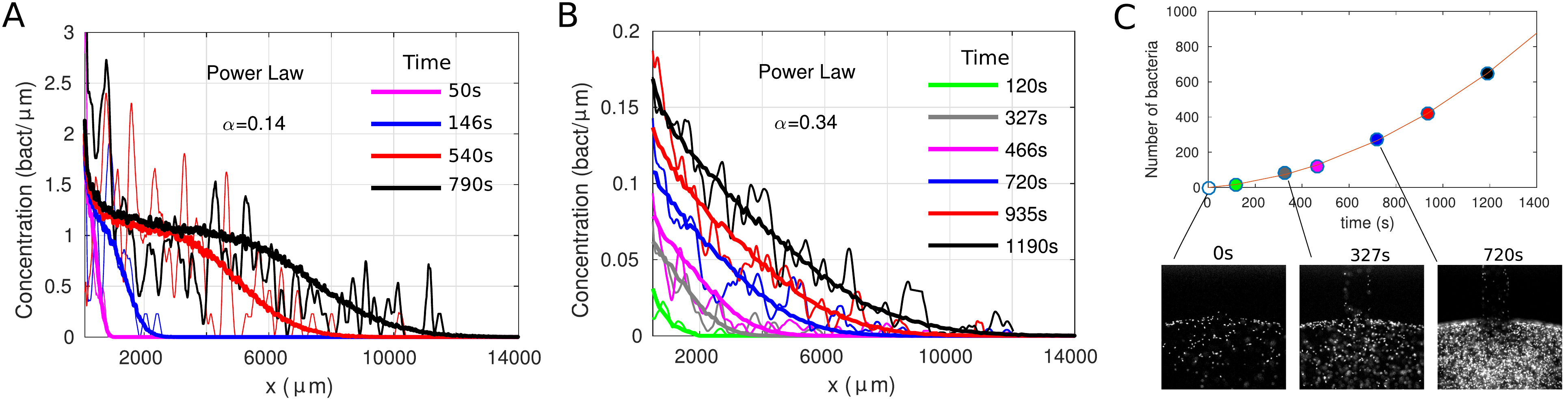}
 \caption{Concentration profiles and boundary conditions. (A) and (B) show the concentration profiles along the channel at different times, for two different flows: $V_\text{f} = \SI{23}{\micro\meter/\second}$  and $V_\text{f} = \SI{80}{\micro\meter/\second}$,  respectively. The average velocity of bacteria is $V_\text{b}=\SI{22}{\micro\meter/\second}$ in both cases. The thicker continuous lines are simulations based on a broad statistical distribution of runs. (C) Number of bacteria inside the channel for the set of scans of panel (B), using the same color code. The three snapshots show the increase of the concentration in the outlet (stock of bacteria) as time goes, therefore, increasing the flux of bacteria into the channel.}
 \label{fig:ExperimentalResults}
\end{figure*}

On Fig.~\ref{fig:pioneers}(A) we show, for two experiments at different flow velocities $V_\text{f}$, the scanning position as a function of time, until the pioneer bacteria are found. The lines topping the scanning positions highlight the advance of pioneers. Notably, these are straight lines [Fig.~\ref{fig:pioneers}(B)], indicating that the contamination front advances at constant speed, $V_\text{cont}$. With the scanning technique, it is not possible to detect whether the pioneering bacteria remain the same, or there is a group of pioneers which alternate positions. However, the tracks on Fig.~\ref{fig:Trajectories} suggest that the second scenario is more plausible. The simulations of the model described later also support that different bacteria take the pioneer role at different moments.

Different perfusion flows and bacterial velocities result in different contamination velocities, as shown in Fig.~\ref{fig:pioneers}(C). In the experiment where  $V_\text{f}=\SI{65}{\micro\meter/\second}$  (average shear rate $\SI{11.0}{\second\tothe{-1}}$) bacteria advance upstream at the amazing velocity $V_\text{cont}=\SI{39}{\micro\meter/\second}$  (\SI{2.3}{\milli\meter/\minute}). In the case  $V_\text{f}=\SI{80}{\micro\meter/\second}$  (nearly four times the average bacterial velocity, average shear rate $\SI{13.6}{\second\tothe{-1}}$), the pioneer bacterial front advances at nearly $V_\text{cont}=\SI{15}{\micro\meter/\second}$  (\SI{0.9}{\milli\meter/\minute}).

The advance of the pioneer contaminants depends on the dispersion of the velocity distribution of the bacteria. The pioneers are likely to be the fastest and more persistent among the population, whereas the upstream contamination is a phenomenon involving the whole bacterial population. To gain insight into the phenomenon, we analyze the concentration profiles that can be reconstructed from scans (see Methods for details).
Figures \ref{fig:ExperimentalResults}(A) and (B) illustrate these profiles for two different experimental conditions. 
As in the case of the most persistent upstream swimmers, we can see that a significant fraction of bacteria are able to move upstream at velocities comparable to the average velocity in the reservoir $V_\text{b}$ [see Fig.~\ref{fig:pioneers}(B)], providing quantitative arguments in favor of the concept of upstream ``super-contamination''.

\section*{Modeling upstream super-contamination} 
\label{sec.model}

\begin{figure*}[t]
\centering \includegraphics[width=1\linewidth]{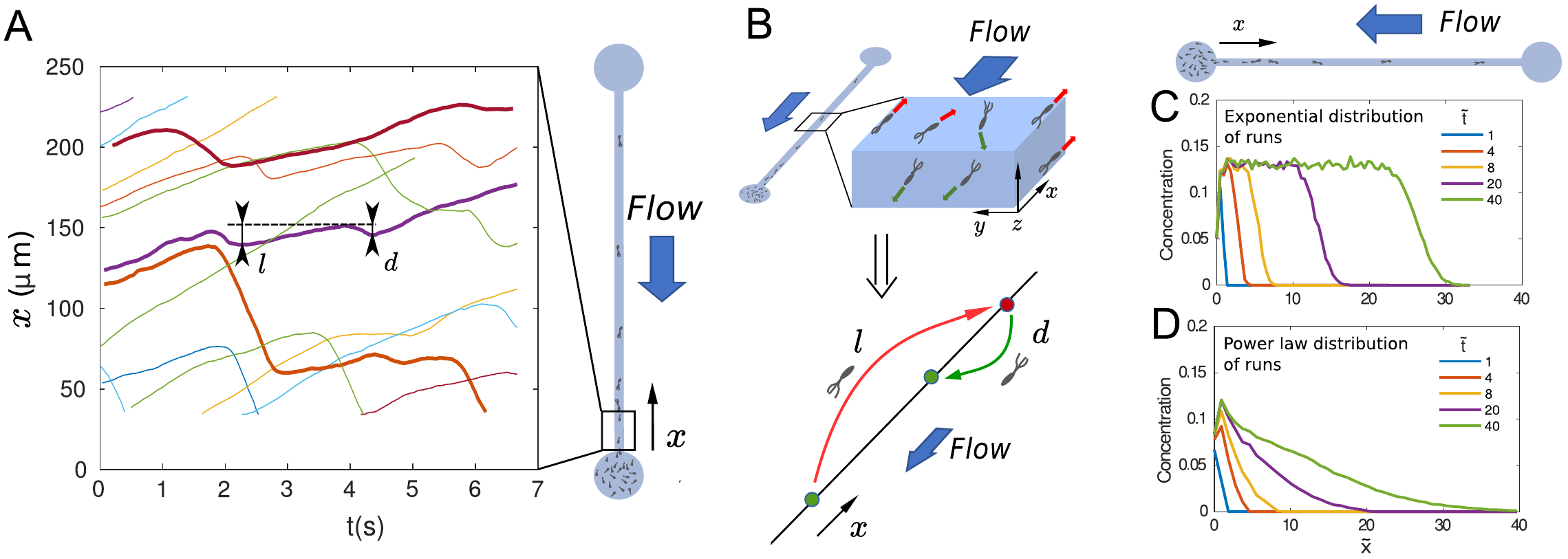}
 \caption{Modelling ``super-contamination". (A) Experimental spatial-temporal plot illustrating the two characteristic lengths associated to the upstream bacterial motion. (B)~Sketch showing the mechanism of upstream contamination in 3D (upper panel) and in the 1D version supporting our biased random walk model (lower panel). (C) and (D) Contamination profiles at different times generated by our model based on Poissonian and broad distributions of run times, respectively. The flux of bacteria from the reservoir was kept constant in the simulations. The parameters for the two distributions assure similar average run times.}
 \label{fig:Experiment_simulation}
\end{figure*}

Although all the physical magnitudes and expressions are defined through the text, for more clarity, Table \ref{Tab:1} shows a compilation of the definitions here used. 

Now we analyze the microscopic mechanisms responsible for the upstream super-contamination. At low flow, the process takes place due to the upstream swimming of bacteria along  edges and surfaces, and only along the edges for higher fluid velocities \cite{Figueroa2015}. When bacteria leave the wall interceptions or the surfaces they are transported downstream. These advected bacteria eventually reattach on the surfaces and again start their upstream migration as in Fig.~\ref{fig:Trajectories} \cite{Hill2007,Altshuler2013,Figueroa2015}.
This upstream-downstream dynamics is shown in supplementary Movie S2. Fig.~\ref{fig:Experiment_simulation}(A) shows the experimental spatial-temporal diagram taken from video V2. We denote $l$ the upstream distance traveled by bacteria between successive detachments and $d$, the downstream transport distance. The distribution of $d$ is difficult to evaluate quantitatively, because for many bacteria the detachment or reattachment locations are not within the window of visualization. However, for the flow of Fig.~\ref{fig:Experiment_simulation}(A), we could evaluate $d$ to be of the order of ten micrometers and it should increase with $V_\text{f}$. Very high fluid velocities will lead to high values of $d$ (bacteria transported farther downstream) and small values of $l$ (bacteria more likely to abandon the surfaces due to high shear), making the contamination impossible when $d > l$.

We propose to model the upstream transport process using a simple one-dimensional biased random walk. A given bacterium performs an upstream displacement $l$, until it detaches from the wall. Then, it will be transported downstream a distance $d$, until it reattaches to the surface, starting its upstream swimming again. Figure \ref{fig:Experiment_simulation}(B) presents a diagram synthesizing these ideas. Our key hypothesis is that the passage from upstream to downstream motion is controlled by the occurrence of tumbling events, where bacteria cease swimming and de-bundle their flagella. Since the attraction to the surfaces and borders is a result of the hydrodynamic interactions of swimmers~\cite{Berke2008,doi:10.1063/1.3676245} and not taking place for passive suspensions~\cite{guazzelli2011physical}, a tumbling bacterium would likely lose  attraction to the surface and be carried downstream. In this picture, the contamination process should depend critically on the detailed statistics of the run and tumble events. 

Under flow, the detachment probability was found to increase leading to a shear-mediated erosion as observed in \citep{Figueroa2015}. 
However, the characteristic shear rates for erosion from surfaces and edges are rather high (\SI{140}{\second\tothe{-1}} and \SI{250}{\second\tothe{-1}}, respectively) \cite{Figueroa2015}.
It is then reasonable to consider that at the low shear rates used in our experiments, desorption is dominated by tumbling events.
From this, we can estimate the upstream swimming distance $l$ as the bacterium velocity $V_\text{b}$ times a characteristic time $\tau$ related to the mean run time. We introduce a parameter $p_e$ equal to the probability of a tumble to be effective, i.e., to produce desorption from surfaces. This leads to $l = V_\text{b} {\tau_\text{run}}/{p_e}$, with $\tau_\text{run}$ being the average run time. Note that $p_e$ is likely to increase as the flow velocity increases.

In the pioneering work of Korobkova \al\cite{korobkova2004molecular}, the motor switching statistics were measured for individual bacteria. The duration of the counter-clockwise state of the motors (which is related to the run mode of bacteria) were found to be largely distributed, as opposed to a simple Poisson process often put forward to describe bacterial motility \citep{berg1972chemotaxis}. Based on this, we approximate the run time statistics using the motor switching statistics by Korobkova \al.

To decipher the role of the run time distribution on the contamination process we  present in our analysis two parallel approaches, one using a Poisson distribution $\psi_\text{P}$ of run times \cite{berg1972chemotaxis}, and a second using a broad distribution illustrated by a power law $\psi_\text{PL}$, from the single flagellum statistics \cite{korobkova2004molecular}.
The corresponding probability distribution functions for run times in the Poisson case is
$ \psi_\text{P}(t) = e^{-t/\tau_\text{P}}/\tau_\text{P}$, 
for an average run time $\tau_\text{run} = \tau_\text{P}$.
 To model the broad distribution we  take
$ \psi_\text{PL}(t) = \gamma/[\tau_0(1+t/\tau_0)^{\gamma+1}]$,
with $\tau_\text{run} = \tau_0/(\gamma -1)$.

We define the dimensionless parameter  quantifying the contamination,
$ \alpha = {d}/{l} = {p_e d}/{(V_\text{b} \tau_\text{run})}$.
When $\alpha \ll 1$, the persistent upstream motion dominates and super-contamination takes place. On the other hand, when  $\alpha \sim 1$, the contamination will be slow since bacteria will be transported downstream almost as much as they can swim upstream between successive detachments. For $\alpha > 1$, no contamination occurs. 

We simulate the evolution of individual bacteria undergoing upstream and downstream transport according to the hypothesis of our model. Detachments from the walls are simulated with a Monte Carlo dynamics following either a power law or Poisson statistics for the run times. Trajectories of bacteria are recorded and accumulated, from which we extract the simulated concentration profiles. 
The resulting contamination profiles are dramatically different for the two distributions. In the Poisson case [Fig.~\ref{fig:Experiment_simulation}(C)], steep fronts move upstream. In the power law case [panel (D)], clear upstream tails composed of very persistent swimmers determine the super-contamination process. 

We can then quantitatively compare the experimental results to the model outcome. One feature to notice from the evolution of the concentrations in Fig.~\ref{fig:ExperimentalResults}(A) and (B) is the increase of bacterial concentration in the reservoir as time increases. This is shown in the snapshots in panel (C) for the set of scans in (B). These bacteria come swimming upstream through the outlet tube and accumulate at the the channel outlet. This effect imposes non-steady boundary conditions at the outlet, which should be explicitly taken into account in the quantitative evaluation of the model. 
To address this issue, we count the total number of bacteria inside the channel for every scan $N_\text{b}$, shown in  the plot of Fig.~\ref{fig:ExperimentalResults} (C).The flux of bacteria into the channel is simply $dN_\text{b}/dt$, which is fitted by a quadratic function and used as the non-steady boundary condition in the simulations.
To identify the contamination parameter $\alpha$ best describing the experiment at a given flow rate, we minimize the squared distance per unit length between the experimental and the simulated profiles. We take into account all the available concentration curves at different times for every set of scans. 

For the broad distribution $\psi_\text{PL}(t)$, we obtain excellent agreement with the experimental contamination profiles, both in space and time. This is shown with thick continuous lines in Fig.~\ref{fig:ExperimentalResults}(A) and (B). The parameters used were $\tau_0 = \SI{1}{\second}$ and $\gamma = 1.2$, corresponding to the measurements on individual flagellar rotation from Korobkova et al. \citep{korobkova2004molecular}.
We can now compare the quantitative results of Fig.~\ref{fig:ExperimentalResults} (A) and (B), which happen to have the same average velocity of bacteria ($V_\text{b}=\SI{22}{\micro\meter/\second}$). The optimal parameter $\alpha$ is bigger in panel (B), as we expected for an experiment with higher fluid velocity, since both $p_e$ and $d$ should be bigger. On the other hand, when using the Poisson process, there was a qualitative disagreement between the model and the experiments, as shown in the SI.

\begin{figure}[h]
\centering \includegraphics[width=\linewidth]{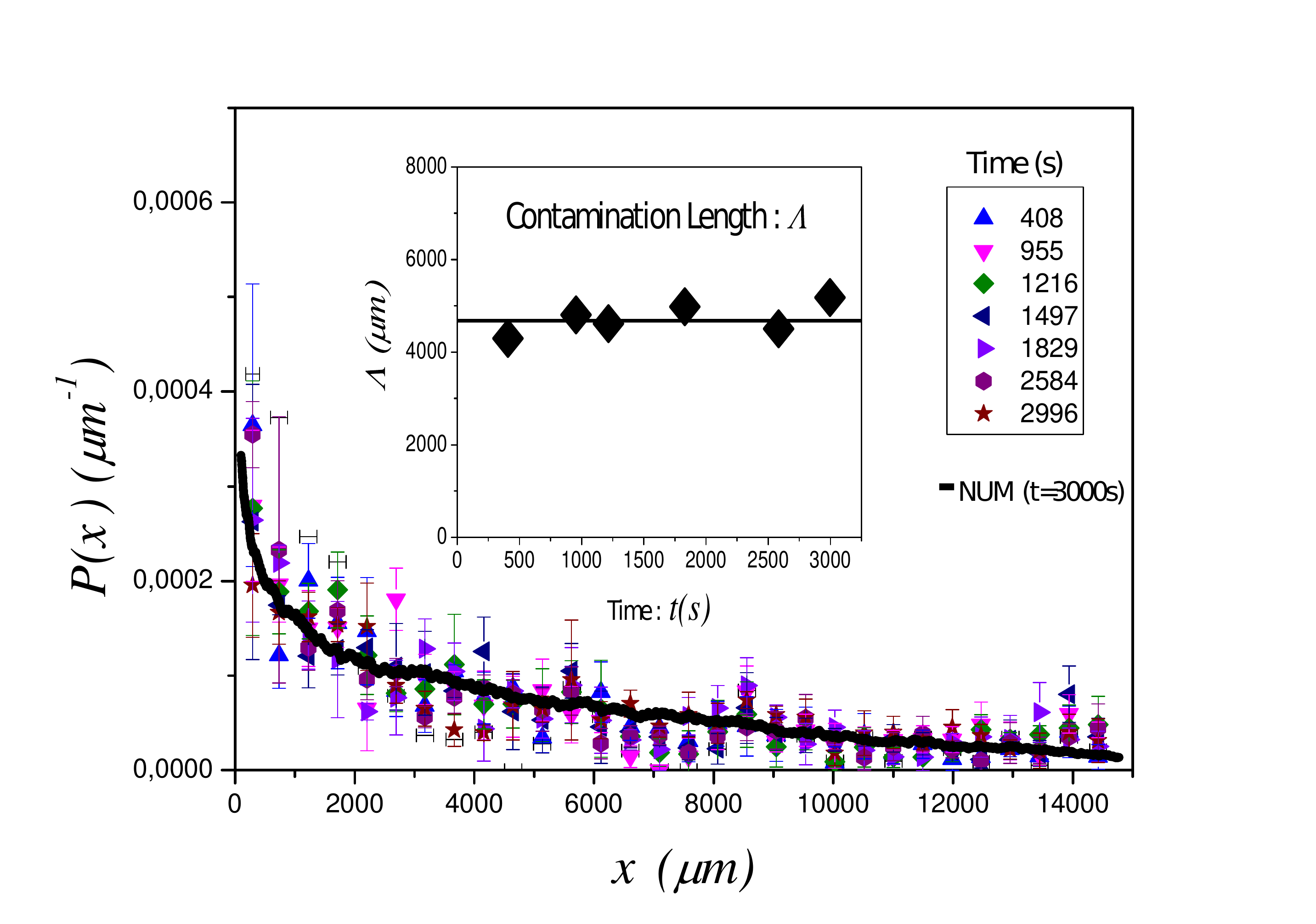}
 \caption{Channel contamination in absence of a flow. Normalized spatial concentration $P(x,t)\equiv c(x,t)/\int c(y,t)dy$  of the bacteria in the channel, for time spanning between \SI{400}{\second} and \SI{3000}{\second}, where $c(x,t)$ are the concentration profiles. The black line is  the result of a numerical random walk model using the switch time distribution $\psi_\text{PL}(t)$. Inset: contamination lengths, $\Lambda(t) = \int x P(x,t) dx$,  for the experiment and the numerical model at $t=\SI{3000}{\second}$.}
 \label{fig:zero_flow}
\end{figure}

For consistency, we now question the results of the contamination process in the absence of flow. The contamination profiles remains essentially localized near the outlet (see Fig.~\ref{fig:zero_flow}). The observed stationary profile stems from the balance between the invasion from the outlet reservoir and the probability to leave the channel after a while by one of the two opposite reservoirs.
As soon as flow is turned on, the situation changes drastically: 
bacteria reorient towards the upstream inlet, leading to the super-contamination process.
At zero flow, the first moment of the distribution characterizes a typical penetration length over a distance $\Lambda \approx \SI{5}{\milli\meter}$  after few minutes (inset Fig.~\ref{fig:zero_flow}).
We simulated a 1D random walk with a probability to change the swimming directions simply triggered by the broad distribution $\psi_\text{PL}(t)$.
For a constant contamination flux of bacteria at the outlet, using a swimming velocity $V_\text{b}=\SI{20}{\micro\meter/\second}$ for a channel of dimension $L = \SI{15}{\milli\meter}$ we obtain, without any adjustable parameter, a  distribution reaching a stationary profile, once again in quantitative agreement with the experimental measurements (see Fig.~\ref{fig:zero_flow}).

\begin{table}
\begin{center}
\begin{tabular}{l l}
\hline \hline \\
$V_\text{b}$ &  Bacterial velocity in the reservoir \\ \\
$V_\text{f}$ &  Maximal flow velocity \\ \\
$V_\text{cont}$ &  Contamination (pioneers') velocity \\ \\
$V_\text{s}=\SI{150}{\micro\meter/\second}$ & Scanning velocity \\ \\
$\tau_\text{run}$ & Average run time \\  \\
$p_e$  &  Probability of a tumble to be effective \\ \\
$l \sim V_\text{b} {\tau_\text{run}}/{p_e}$ &  Upstream swimming distance \\ \\
$d$ &  Downstream transport distance \\ \\
$\alpha=\frac{d}{l} $ & Contamination parameter \\ \\
$ \psi_\text{P}(t) = \frac{e^{-t/\tau_\text{P}}}{\tau_\text{P}}$ &  PDF in the Poisson case \\ 
 &  $\tau_\text{run} = \tau_\text{P}$ \\ \\
$ \psi_\text{PL}(t) = \frac{\gamma}{\tau_0(1+t/\tau_0)^{\gamma+1}}$ &   PDF in the power law case \\
  & $\tau_0 = \SI{1}{\second}$, $\gamma = 1.2$, $\tau_\text{run} = \tau_0/(\gamma -1)$ \\ \\
$N_\text{b}$ &  Number of bacteria inside the channel \\ \\
$c(x,t)$ & Concentration of bacteria \\ & in the channel \\ \\
$P(x,t)= \frac{c(x,t)}{\int c(y,t)dy}$ & Normalized spatial concentration \\ & of bacteria in the channel \\ \\
$\Lambda(t) = \int x P(x,t) dx$ & Contamination length  \\ \\
\hline
\end{tabular}
\end{center}
\caption{Summary of physical magnitudes.}
\label{Tab:1}
\end{table}

\section*{Conclusions}
 
We showed in a simple microfluidic experiment that bacteria can rapidly contaminate initially clean environments by propagating upstream in a narrow channel, over long distances and for a significant range of flow rates. The bacteria pioneering the contamination advance at constant velocity for distances over 1 centimeter. The contamination process results in concentration profiles with long tails that we characterize in space and time. 

Solving numerically a simple one-dimensional model of a biased random-walk, we relate quantitatively, the spatio-temporal contamination profile to a broad distribution of run times stemming from the clockwise/counterclockwise statistics of the bacterial motor rotation. To our knowledge, this is the first time that a quantitative relation is made between single cell experiments on bacterial motors and the outcome of a macroscopic transport process. This puts forward that singular features of the \rt~ statistics, borne in the stochasticity in the chemotactic circuitry, have a definite influence on macroscopic transport, in agreement with recent observations from 3D Lagrangian tracking~\cite{Figueroa_Thesis, figueroa20183d}. 

In practice, our measurements suggest that swimming bacteria can overcome distances comparable to the sizes of animal organs (tens of centimeters) in some tens of minutes or a few hours.
As an example, ureters in the human urinary tract are a possible scenario for super-contamination. These are tubes with muscular walls that undergo successive waves of active muscular contraction, to move liquid from kidney to bladder. When totally contracted, ureters collapse to slit-shaped, very confined cross-sections, possibly favorable to upstream bacterial migration. When fully distended, we estimate shear rates of around 10--\SI{60}{\second\tothe{-1}}~\cite{Ureter1989flow}. At these low shear rates, bacteria undergo little erosion from surfaces and edges~\cite{Figueroa2015}. Contamination fronts advancing at 15--\SI{25}{\micro\meter/\second} could overcome the length of the ureters (200--\SI{300}{\milli\meter}) and travel from the bladder to the kidneys in 3 to 7 hours, possibly starting a renal infection. 

The super-contamination could be relevant in other scenarios: 
Histological studies of the bovine cervical mucosa showed longitudinal grooves of cervical folds, which maintained continuity throughout the cervix \cite{june1989study}. These geometrical conditions potentially facilitate the fast upstream migration of bacteria with a subsequent infection.
Acute cholangitis, another medical emergency, is usually caused by bacteria ascending from the duodenum through the bile duct and infecting it \cite{kinney2007management}, especially when it is partially obstructed and therefore, provides a very confined environment ideal for upstream contamination.

\section*{Methods}

\subsection*{Bacterial strains and culture}
We use RP437 \textit{E. coli} bacteria. The cells are cultured overnight at \SI{30}{\degreeCelsius} in M9 minimal medium supplemented with \SI{1}{\milli\gram/\milli\liter} casamino acids and \SI{4}{\milli\gram/\milli\liter}  glucose. Next, bacteria are washed twice by centrifugation ($2300g$  for \SI{5}{\minute}) and the cells are re-suspended into a motility medium containing $10$ mM potassium phosphate pH $7.0$, $0.1$ mM K-EDTA, $34$ mM K-acetate, $20$ mM sodium-lactate and $0.005\ \%$ polyvinylpyrrolidone (PVP-40). In this medium, bacteria are able to live and swim but do not divide. 

\subsection*{Microfluidic device and procedure}

The experimental cell is a microfluidic channel (rectangular cross-section, width $w = \SI{40}{\micro\meter}$, height $h = \SI{11}{\micro\meter}$, length \SI{15}{\milli\meter}) ending it two cylindrical capacities. It is made in PDMS using a conventional soft photolithography technique, and assembled onto glass plates previously coated with a thin layer of PDMS. 
Stainless steel tubes of  \SI{1}{\milli\meter} diameter were inserted at each end of the channel in the cylindrical capacities, connected to large liquid reservoirs through plastic flexible tubes. After perfectly filling the microfluidic system with the stock solution without bacteria, the metallic connector from the outlet was replaced by a similar one connected to a big reservoir containing the same liquid as the inlet, plus bacteria.  As a result, we start the experiment with a bacterial suspension located at the left end of the channel (see the panel corresponding to $t=0$ in Fig.~\ref{fig:scan_montage_vertical}), while the rest of its length was filled with a bacteria-free medium.

The system was placed on an inverted microscope (Zeiss-Observer, Z1) with an $xy$ mechanically controllable stage from ASI, a digital camera ANDOR iXon 897 EMCCD ($512\times\SI{512}{pix\tothe2}$ at a frequency of $f = \SI{30}{fps}$) with a $40\times$ magnification objective.

Flow is established by imposing a small height difference between the reservoirs, which allows us to work with very small flow rates.
We visualize all the bacteria along the microchannel height. As time passes, bacteria migrate upstream along the channel. A single realization of the contamination experiment consists of periodically scanning the channel, to count bacteria along its length. To do so, we move the microscope stage along the channel axis at a scanning velocity of $V_\text{s} = \SI{150}{\micro\meter/\second}$ while recording a video. Later on, on image post-processing, we relate the number of bacteria in each frame to its distance $x$ from the reservoir.

Between subsequent scans we take a video at a fixed position using direct light, enabling the visualization of tracers. 
The velocity profile was determined for each applied pressure difference by tracking the plastic beads. 


\subsection*{Construction of the bacterial concentration profiles along the channels}


To obtain the contamination profiles from the analysis of the scans, we count the number of bacteria in each frame of the scans. 
Since the distance between two consecutive pictures, $\Delta x_\text{s} = V_\text{s}/f= \SI{150}{\micro\meter/\second}/\SI{30}{fps} = \SI{5}{\micro\meter}$, is smaller than the piece of the channel imaged in one frame ($L_x = \SI{160}{\micro\meter}$), some bacteria are imaged several times. To obtain a concentration profile we normalize the total number of bacteria detected by the average number of times that a bacterium was recorded: $L_x / \Delta x_\text{s}$. As the profiles do not come from snapshots, but from scans, they are stretched, showing tails longer than what they really are. We correct for the stretching as follows:
Consider a bacterium in the position $x > 0$ at the starting moment of the scan and swimming upstream with a speed $V_\text{b}$. This bacterium will be registered at the moment $t > 0$ when it has traveled a distance $\Delta x = V_\text{b} t$. In the reference frame of the channel, the objective, initially at $x=0$, would have traveled a distance $x_m = x + \Delta x = V_\text{s} t$ when it captures the bacterium. Here $V_\text{s}$ is the scanning velocity ($V_\text{s} > V_\text{b}$). From the equality of times we obtain $\frac{\Delta x}{x_m} = \frac{V_\text{b}}{V_\text{s} }$, which shows that the deformation is linear with the distance to the reservoir. In our experiments the coefficient is in the range $0.02 < \frac{\Delta x}{x_m} < 0.2$. With this principle we re-scaled the $x$ axis to reduce the profile stretching.
The new $x$ values are $x = x_m - \Delta x = x_m \left(1 - \frac{V_\text{b}}{V_\text{s}} \right)$, where $x_m$ is the measured coordinate from the scan.

\section*{Acknowledgements}
We acknowledge A. Rousselet and R. Garc\'ia for useful discussions. 
N.F.M. thanks support by the Pierre-Gilles de Gennes Foundation. 
E.A. and A.R. acknowledge ``Joliot Curie'' Chairs. 
We acknowledge the financial support of the ANR 2015 ``Bacflow'' and the Franco-Chilean EcosSud Collaborative Program C16E03. 
A.L. and N.F.M. acknowledge support from the ERC Consolidator Grant PaDyFlow under grant agreement 682367. 
R.S. acknowledges the Fondecyt Grant No.\ 1180791 and Millenium Nucleus Physics of Active Matter of the Millenium Scientific Initiative of the Ministry of Economy, Development and Tourism (Chile).

%
%

%

\section*{Supplementary Information}

\subsubsection*{Dimensionless parameter}

\begin{figure}[h]
\centering      \includegraphics[width=0.7\linewidth]{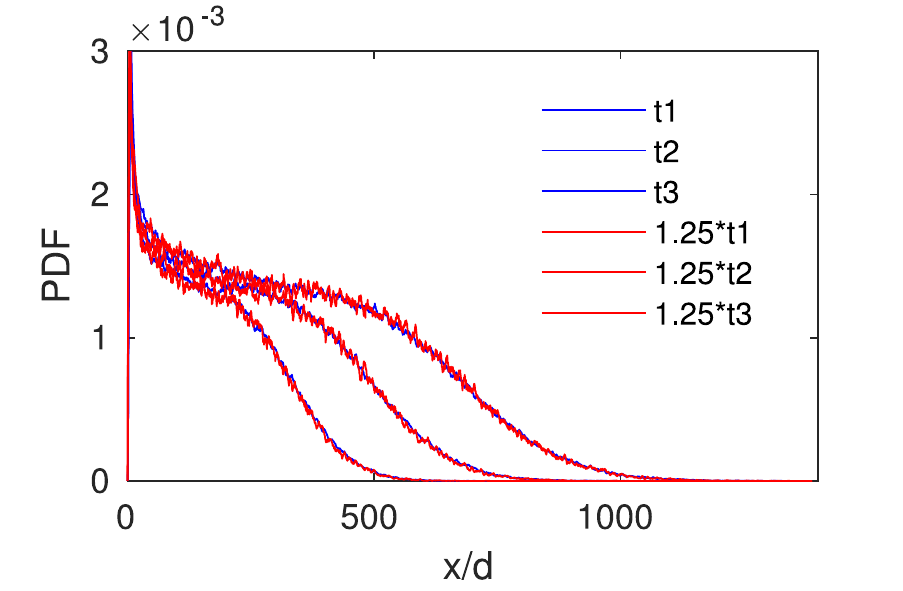}
 \caption{Simulations with different $p_e$ and $d$, but similar $\alpha$. After the proper normalization both sets of simulations superpose, showing that $\alpha$ is the dimensionless parameter in the problem. (pe = 1, td = 1.2; pe = 0.8, td = 1.5)}
 \label{fig:same_alpha}
\end{figure}

To verify that $\alpha$ is the only dimensionless parameter in the simulations, we run simulations using two different values of $p_e$ and $d$, leading to the same $\alpha$. After normalization in each case, by the characteristic time of the problem $\tau / p_e$ and the dimension $d$, both sets of concentration profiles superpose, as shown in Fig. \ref{fig:same_alpha}.

\subsubsection*{Simulations with Poisson statistics}

\begin{figure}
\centering      \includegraphics[width=0.7\linewidth]{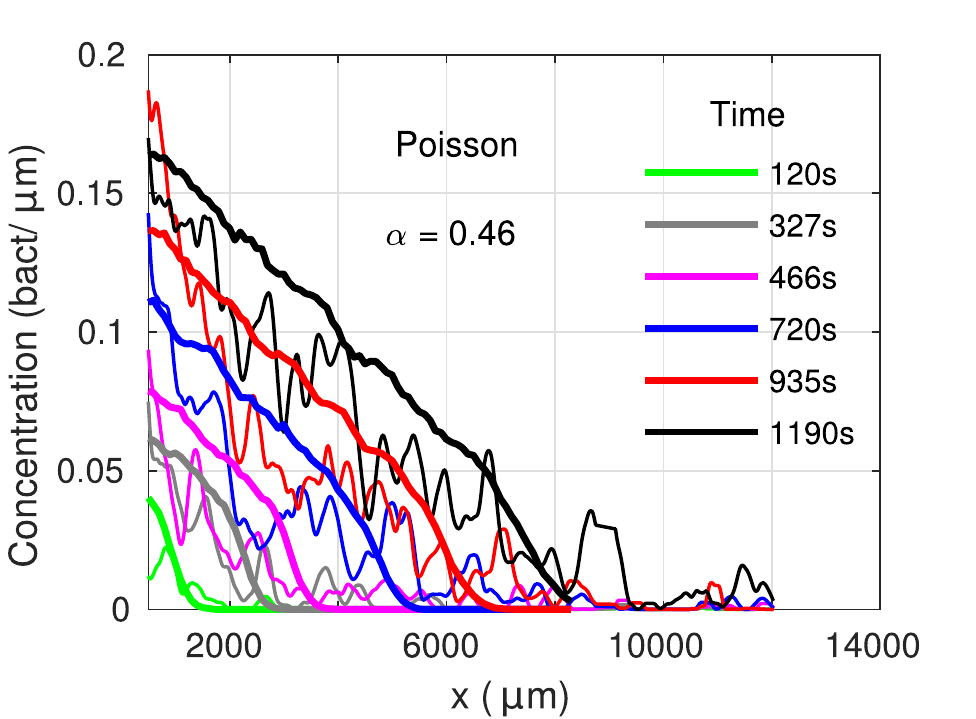}
  \caption{Superposition of simulations and contamination experiments using the Poisson statistics for the indicated $\alpha$. The thin lines are the experimental concentrations and the thick ones of same color their respectively simulated curves. There is a poor agreement between the experimental and simulated curves, evidencing the failure of the Poisson process for modeling the desorption. }
 \label{fig:Poisson_80umps}
\end{figure}

Figure \ref{fig:Poisson_80umps} shows the superposition of contamination experiments and simulations using the Poisson process for the desorption statistics. The qualitative agreement is poor. 

\subsubsection*{Comparison between distributions}

For the quantitative comparison we will search to minimize the squared distance per unit length between the experimental profile and the simulated one for a given value of $\alpha$. We defined as 
\begin{eqnarray*}
 F(\alpha)  =  \sum_i^{N_{curves}} \frac{1}{N_{bins } \Delta x} \sum_j^{N_{bins}} \left( N_{Real} (x_j, t_i) \right. \\
 \;\ \left. - N_{Sim}^{\alpha} (x_j, t_i)  \right)^2 \mbox{,}
\end{eqnarray*}
where $i$ counts the number of profiles corresponding to different scanning times and $j$ counts the bins along the channel. $N_{Real}$ and $N_{Sim}$ are the numbers of particles in experiment and simulation respectively, corresponding to the bin of width $\Delta x$ located at $x_j$. 

\begin{figure}
\centering      \includegraphics[width=0.7\linewidth]{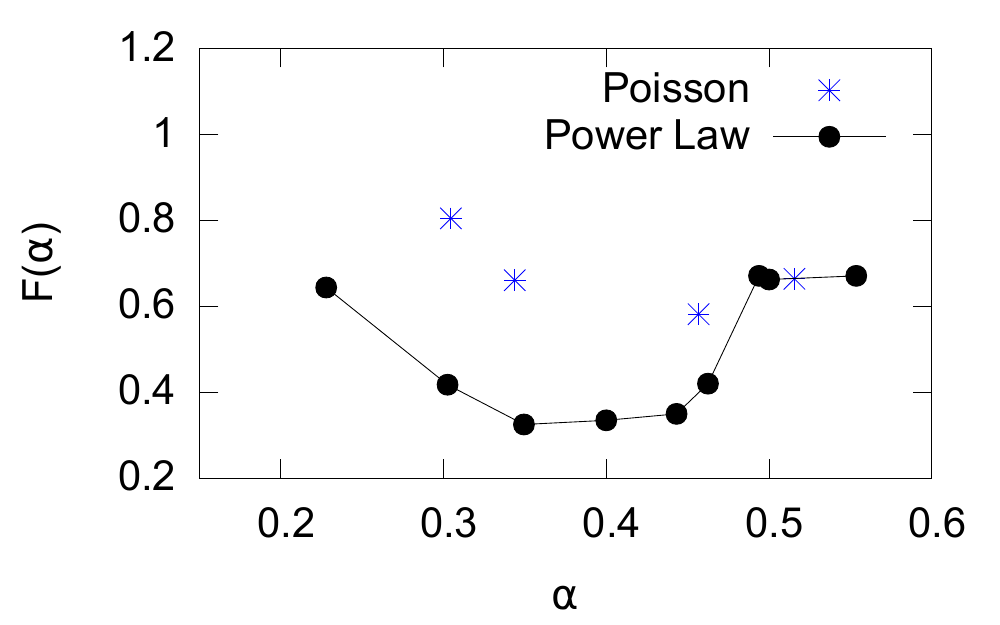}
  \caption{Optimization factor as a function of the dimensionless parameter ($\alpha$) of the model for the Poissonian and power law run time distributions, corresponding to an experimental case with average flow velocity $V_\text{f} = 80 \mu m/s$. There exists a clear minimum for the power law distribution, while no optimization seems to take place for the exponential law.}
 \label{fig:f_80umps}
\end{figure}

When exploring the parameter $\alpha$ using the Poissonian law for the run distribution, we do not find an evident minimum of the function $F(\alpha)$. However, for the power law distribution function a clear minimum exists and its value is lower than the lower ones found in the exponential case. The two curves are shown in Fig. \ref{fig:f_80umps}.

\paragraph*{Video 1} shows a sequence of three scans starting at times $0s$, $327s$ and $719s$ for a contamination experiment with maximum flow velocity $80 \mu m/s$. The corresponding concentration profiles are those of Fig. 4 (B) (main article) and Fig. \ref{fig:Poisson_80umps}.

\paragraph*{Video 2} was taken from a fixed reference frame with respect to the microfluidic channel. It shows the bacterial upstream-downstream dynamics for a contamination experiment with maximum flow velocity $80 \mu m/s$.

\end{document}